\documentclass[twocolumn,pra]{revtex4}

\begin{document}

\title{Comment on ``New Methods for Old Coulomb Few-Body
Problems''}
\author{V.~S.~Zotev}
\affiliation{Los Alamos National Laboratory, Biophysics Group, MS
D454, Los Alamos, NM 87545, USA}

\begin{abstract}
In this Comment on the above mentioned paper by F.~E.~Harris,
A.~M.~Frolov, and V.~H.~Smith, we briefly review our contributions
to development of new methods for solution of the Coulomb
four-body problem. We show that our research group, headed by
Prof. T.~K.~Rebane, had a priority in using the fully correlated
exponential basis for variational calculations of four-body
systems. We also draw attention to the fact that our group
subsequently implemented a more advanced method, which uses highly
efficient exponential-trigonometric basis functions for solution
of the same problem.
\end{abstract}

\maketitle

The paper ``New Methods for Old Coulomb Few-Body Problems''
\cite{har04} deals with analytical and methodological issues
involved in the implementation of two methods -- the Hylleraas
method and what the authors call the ``exponential ansatz''
approach. It is stated in the abstract that the second method
``has only recently been implemented for four-body systems''
\cite{har04}. The last section of the paper is entitled
``Preliminary Results'' \cite{har04}.

The main purpose of this Comment is to call attention to our works
in this field and show that the method using a large fully
correlated exponential basis for solution of the Coulomb four-body
problem was in fact implemented and successfully applied by our
group ten years ago.

A general method for analytic evaluation of four-particle
integrals was developed by Fromm and Hill in their well-known
pioneering work \cite{hil87} in 1987. This method makes it
possible to perform direct variational calculations of four-body
systems using exponential basis functions of all six interparticle
separations. The first such calculation of two model systems was
carried out by Rebane and Yusupov in 1992 \cite{reb92}. Rebane
showed in 1993 how the number of different types of four-body
integrals that need to be computed can be reduced from 43 to only
seven \cite{reb93}. The first variational calculation of the
positronium molecule $e^{+}e^{-}e^{+}e^{-}$ in the fully
correlated exponential basis was performed by Rebane, Zotev, and
Yusupov in 1996 \cite{reb96}. It employed 90 symmetrized
exponential functions. This work was followed by calculations of
eight symmetric mesomolecules in the basis of 240 exponentials
\cite{zot98}. In 1998, the original method by Fromm and Hill was
generalized by Zotev and Rebane to allow evaluation of
four-particle integrals with complex exponential parameters. This
enabled us to perform the first variational calculation of the
positronium molecule and three other systems using the basis of
extremely flexible exponential-trigonometric functions
\cite{zot00}. It was shown that a single exponential-trigonometric
function is as efficient as seven exponentials in the case of
$e^{+}e^{-}e^{+}e^{-}$, and is flexible enough to allow direct
variational computation of even adiabatic four-particle systems
such as the hydrogen molecule \cite{zot00}. The
exponential-trigonometric basis was initially introduced by our
group in the three-body problem \cite{reb90}. A detailed
description of our methods using the exponential and
exponential-trigonometric basis functions for direct variational
solution of the Coulomb four-body problem was published in 2002
\cite{zot02}.

Harris, Frolov, and Smith implemented the exponential basis and
reported a calculation of the positronium molecule with \emph{two}
exponential functions in 2003 \cite{har03}. They used a modified
Fromm and Hill formula \cite{har97}, and argued that the advantage
of their approach was ``better formal treatment of termwise
singularities'' \cite{har03}. They wrote in the discussed paper
that their refinements enabled them to find ``a two-configuration
wavefunction with energy -0.513237 hartree, comparable to the
result obtained by Rebane \textit{et al.} using four
configurations'' \cite{har04}.

Our extensive practical experience shows that the problem of
termwise singularities is not serious, because parameters of basis
functions can easily be chosen to avoid singularities in all the
integrals to be computed. Our calculations of the positronium
molecule \cite{reb96} and mesomolecules \cite{zot98} with the
proper parameter selection exhibited perfect numeric stability and
steady convergence.

The authors' comparison of their two-function calculation and our
four-function result is performed incorrectly. In their
calculation \cite{har03}, the two functions were optimized
\emph{simultaneously}. In our work \cite{reb96}, the basis was
expanded gradually, and basis functions were optimized \emph{one
at a time}. Naturally, the simultaneous optimization provides
better results. Moreover, our 1996 work included not just four,
but a total of 90 exponential functions, and yielded a variational
energy value of -0.515920 hartree \cite{reb96}.

Finally, we have to mention the following. Frolov and Smith
\cite{fro01}, Frolov \cite{fro01b}, Frolov and Bailey
\cite{fro03}, when discussing the Fromm and Hill four-body
integrals in their 2001-03 articles, did not reference our works
at all. Similarly, Frolov made no reference to our original works
\cite{reb90} in his 1998-03 papers, which used the
exponential-trigonometric basis in the three-body problem
\cite{fro98}. In the discussed paper \cite{har04}, Harris, Frolov
and Smith cite our 1996 calculation of the positronium molecule
\cite{reb96}, but do not reference any subsequent publications
\cite{zot98, zot00, zot02}, particularly the 2002 article
\cite{zot02}, which provides a detailed review of our methods.
Frolov and Smith follow the same line in another recent paper
\cite{fro04}. The most advanced and accurate method, which uses
highly flexible exponential-trigonometric functions for
calculations of four-particle systems with arbitrary particle
masses \cite{zot00,zot02}, has never been mentioned by Harris,
Frolov, and Smith. Even the latest and most comprehensive review
article by Harris \cite{har04b} makes no explicit mention of this
method. We leave these facts without comment.

\end{document}